\documentclass[final,5p,times,twocolumn]{elsarticle}

\usepackage{amssymb}
\usepackage{txfonts}
\usepackage{url}
\usepackage{lineno}

\usepackage{hyperref}
\usepackage{xcolor}

\usepackage{dblfloatfix} 

\makeatletter
\def\ps@pprintTitle{%
  \let\@oddhead\@empty
  \let\@evenhead\@empty
  \let\@oddfoot\@empty
  \let\@evenfoot\@oddfoot
}
\makeatother

\begin{document}

\begin{frontmatter}

\title{Light Monitoring System for the Lead Tungstate Calorimeter in Hall D at Jefferson Lab  \tnoteref{notice} }

\author[jlab]{A.Somov\corref{cor1}}
\ead{somov@jlab.org}
\author[jlab]{V.Berdnikov}
\author[erphy]{H.Voskanyan}
\author[jlab]{A.Asaturyan}
\author[uncw]{L.Gan}
\author[jlab]{S.Taylor}
\author[jlab]{F.Barbosa}
\author[jlab]{C.Stanislav}
\author[jlab]{V.Popov}
\author[uva]{I.Somov}
\author[jlab]{I.Jaegle}
\author[jlab]{A.Smith}
\author[jlab]{H.Egiyan}
\author[jlab]{B.Bunton}
\address[jlab]{Thomas Jefferson National Accelerator Facility, Newport News, VA 23606, USA}
\address[erphy]{A. I. Alikhanian National Science Laboratory (Yerevan Physics Institute), 0036 Yerevan, Armenia}
\address[uncw]{University of North Carolina at Wilmington, Wilmington, NC 28403, USA}
\address[uva]{University of Virginia, Charlottesville, VA 22904, USA}

\cortext[cor1]{Corresponding author. Tel.: +1 757 269 5553;  fax: +1 757 269 6331.}

\begin{abstract}
A new electromagnetic calorimeter composed of 1596 lead tungstate (PbWO$_4$) scintillating crystals has been constructed for the GlueX detector in Hall D at Jefferson Lab. The calorimeter is equipped with a light monitoring system that uses light-emitting diodes. The light monitoring system was fabricated, installed, and integrated into the GlueX trigger system. It was successfully operated during detector commissioning and data collection. The paper describes the design, installation, and performance of the light monitoring system.

\end{abstract}

\begin{keyword}
Lead tungstate calorimeter, light monitoring
\end{keyword}

\end{frontmatter}

\section{Introduction}

The GlueX detector~\cite{gluex_det}, located in Experimental Hall D at Jefferson Lab (JLab), is a large-acceptance forward magnetic spectrometer designed to conduct experiments using a photon beam incident on various targets. The Jefferson Lab Eta Factory (JEF)~\cite{jef} is one such experiment, whose main physics goal is to study the decays of $\eta^{(\prime)}$ mesons into final states involving photons. This experiment requires good reconstruction of photons in the forward direction, which was achieved by replacing the inner section of the GlueX forward lead-glass calorimeter with high-granularity, high-resolution scintillating crystals.

A new Electromagnetic CALorimeter (ECAL)~\cite{ecal}, based on lead tungstate scintillating crystals, was constructed and successfully commissioned in April 2024. The ECAL comprises an array of $40\times 40$ modules, with a $2\times 2$ module hole at the center for the photon beam. Each module consists of a $2\;\rm{cm} \times 2\;\rm{cm} \times 20\;\rm{cm}$ rectangular scintillating crystal, coupled to a Hamamatsu R4125 photomultiplier tube (PMT) via an acrylic light guide. Signal pulses are digitized using an Analog-to-Digital Converter (ADC), designed at JLab, operated at a sampling rate of 250 MHz~\cite{fadc250}.

To monitor calorimeter performance in real time and assist with PMT gain calibration, we designed and installed a Light Monitoring System (LMS). Light from multiple Light-Emitting Diodes (LED) is distributed to the face of each crystal using optical fibers. The LMS was integrated into the GlueX trigger system and operated continuously during data taking. A similar LMS design was previously tested with a calorimeter prototype used during the PrimEx $\eta$ experiment from 2019 to 2022~\cite{fcal2}.

This article is organized as follows: the design and installation of the ECAL LMS are presented in Section~\ref{sec:lms_design}; the integration of the LMS into the GlueX trigger system is described in Section~\ref{sec:lms_trigger}; and the performance of the LMS is discussed in Section~\ref{sec:lms_performance}.
\begin{figure}[htb]
\begin{center}
\includegraphics[width=1.0\linewidth,angle=0]{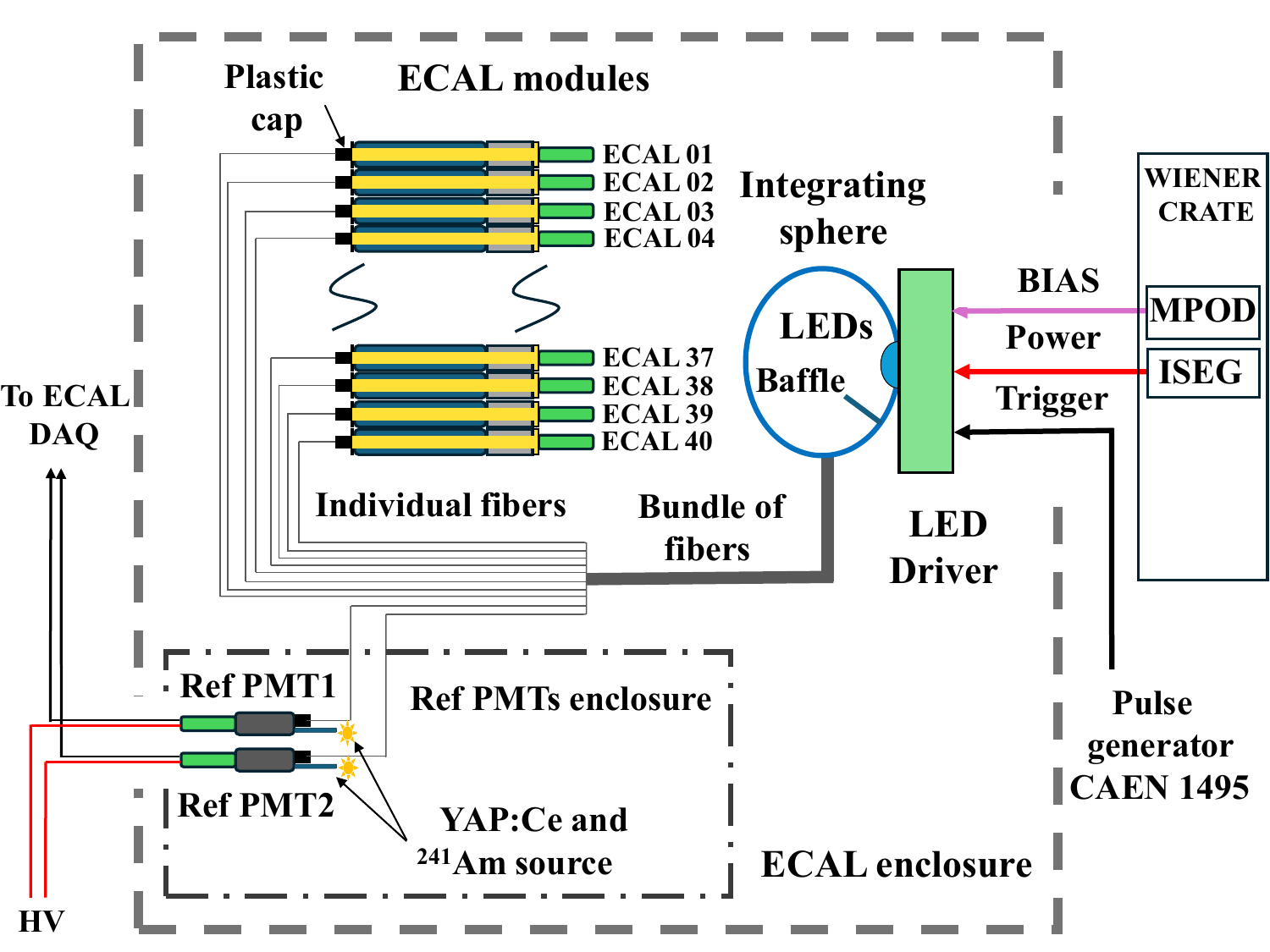}
\end{center}
\caption{Schematic view of the ECAL light monitoring system. }
\label{fig:ecal_lms}
\end{figure}
\begin{figure}[htb]
\begin{center}
\includegraphics[width=1.0\linewidth,angle=0]{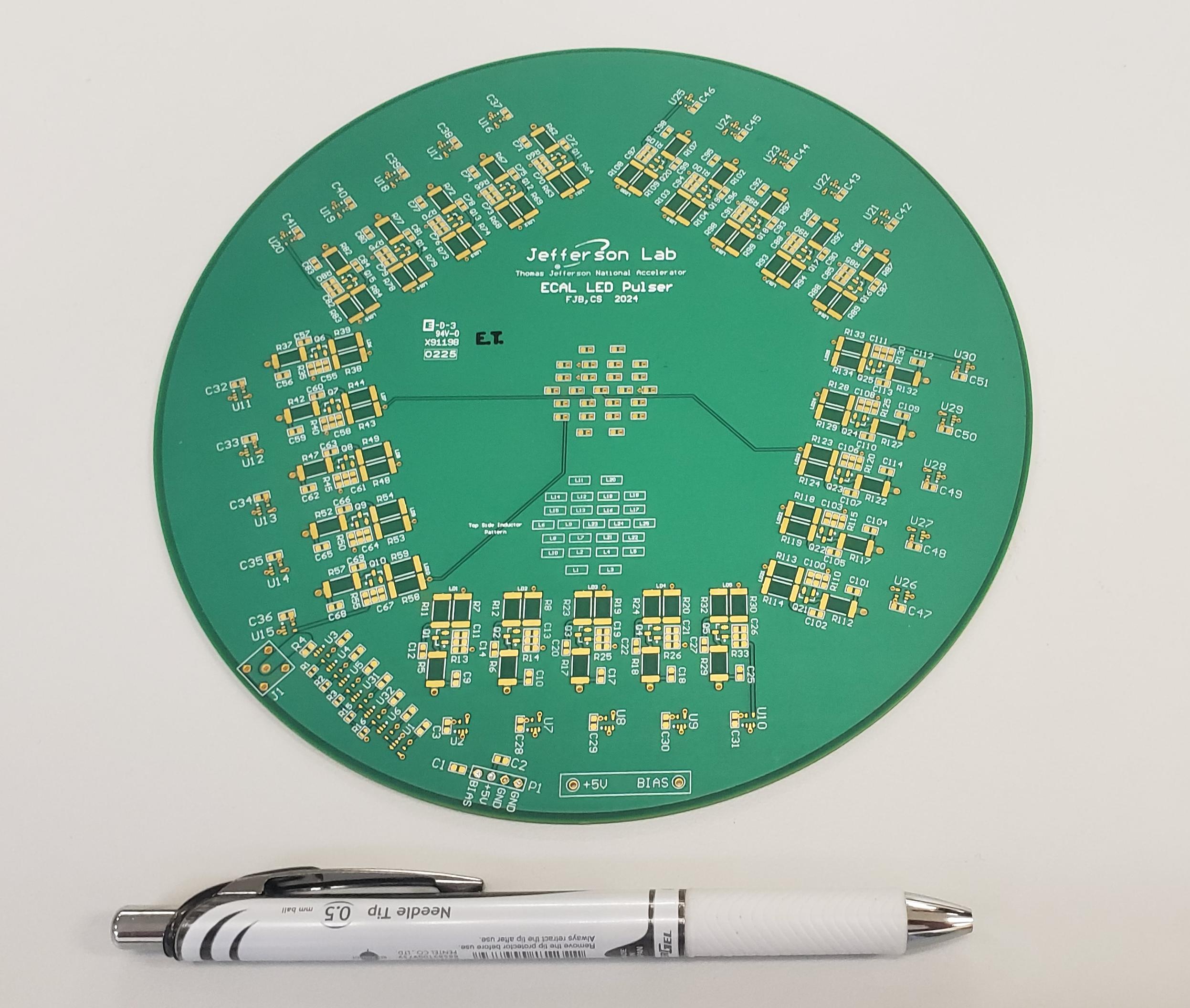}
\end{center}
\caption{The printed circuit board with the light monitoring system driver components and footprints for 25 LEDs in the middle. }
\label{fig:lms_driver_pcb}
\end{figure}

\section{Design and installation of the light monitoring system}
\label{sec:lms_design}
A schematic view of the ECAL LMS is presented in Fig.~\ref{fig:ecal_lms}. The light is produced by multiple LEDs, mounted on an integrating sphere, also known as Ulbricht sphere, to ensure uniform mixing of light. The light is distributed to the face of  each ECAL crystal module via acrylic optical fibers. The LMS system performance parameters, such as light flash intensity and frequency can be remotely controlled. The stability of the LMS is monitored using two reference PMTs, calibrated against an $\alpha$-source-activated light pulse unit that provides a stable light flash amplitude. Both the light source and the reference PMTs are positioned inside a detector dark room, where the ambient temperature is maintained to within better than $1{}^\circ C$ during the data-taking period. Temperature stabilization is critical for the lead tungstate crystals used in the calorimeter, as their light yield is strongly temperature dependent.

\subsection{Light source}
\label{sec:lms_driver}
The light source consists of 25 blue surface-mount LEDs with a central wavelength of 470 nm and a spectral bandwidth of $\Delta \lambda$ = 25 nm. Each LED has a viewing angle of 140° and a physical size of $3.2\;{\rm mm} \times 1.6\; {\rm mm}$. The typical operating voltage is 3.2 V. The LEDs are mounted on a Printed Circuit Board (PCB) along with the LED driver circuit, which is shown in Fig.~\ref{fig:lms_driver_pcb}.  The PCB is mounted onto the input optical port of the integrating sphere, which is supplied by Edmund Optics. The flange of the standard port adapter is glued to the PCB to ensure secure attachment. The diameter of the input port is 25.4 mm, and the diameter of the integrating sphere is 152.4 mm.

\begin{figure*}[!t]
\begin{center}
\includegraphics[width=1.0\linewidth,angle=0]{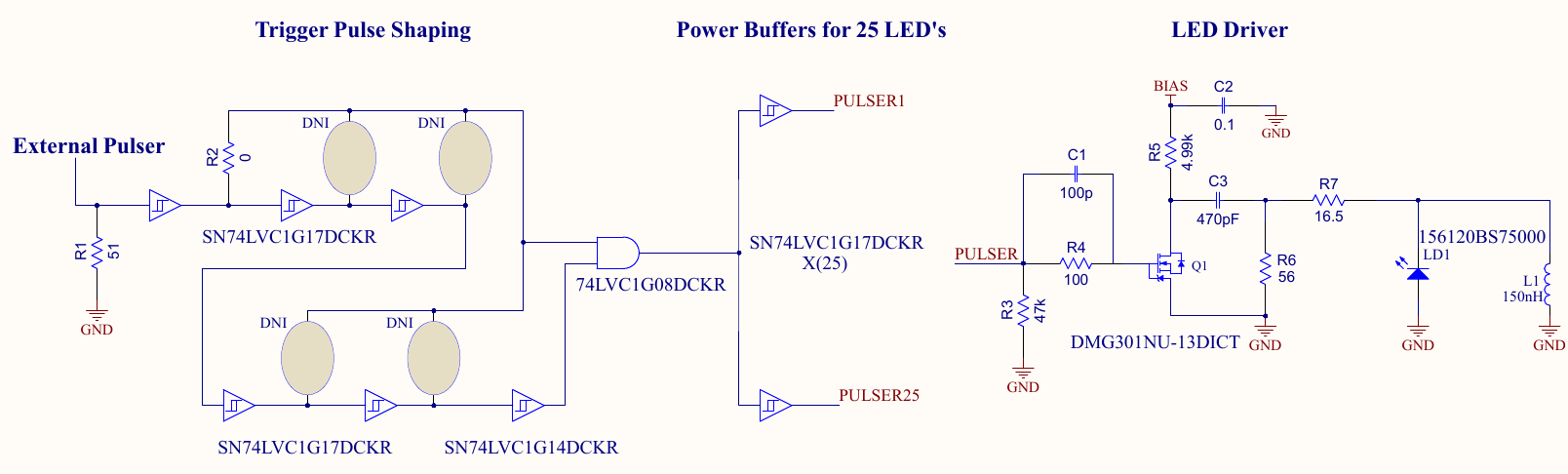}
\end{center}
\caption{Schematic view of the LED driver designed at Jefferson Lab. The main components are the trigger pulse shaper, power buffers for 25 LEDs, and individual LED drivers.}
\label{fig:lms_driver}
\end{figure*}

A schematic of the LED driver is shown in Fig.~\ref{fig:lms_driver}. The LED flash is initiated by an external pulse generator, which sends trigger pulses to the LED driver. The driver operates in two main stages. First, a short trigger pulse with a predefined width is generated from the input signal of the external pulse generator. Then, this pulse is used to activate the driver circuit of each individual LED. Specifically, it switches on a Metal-Oxide-Semiconductor Field-Effect Transistor (MOSFET), allowing the stored charge in a capacitor to be discharged through the LED, producing  a high-intensity flash. 

\begin{figure*} [!b]
\begin{center}
\begin{tabular}{c}
\includegraphics[width=0.5\linewidth]{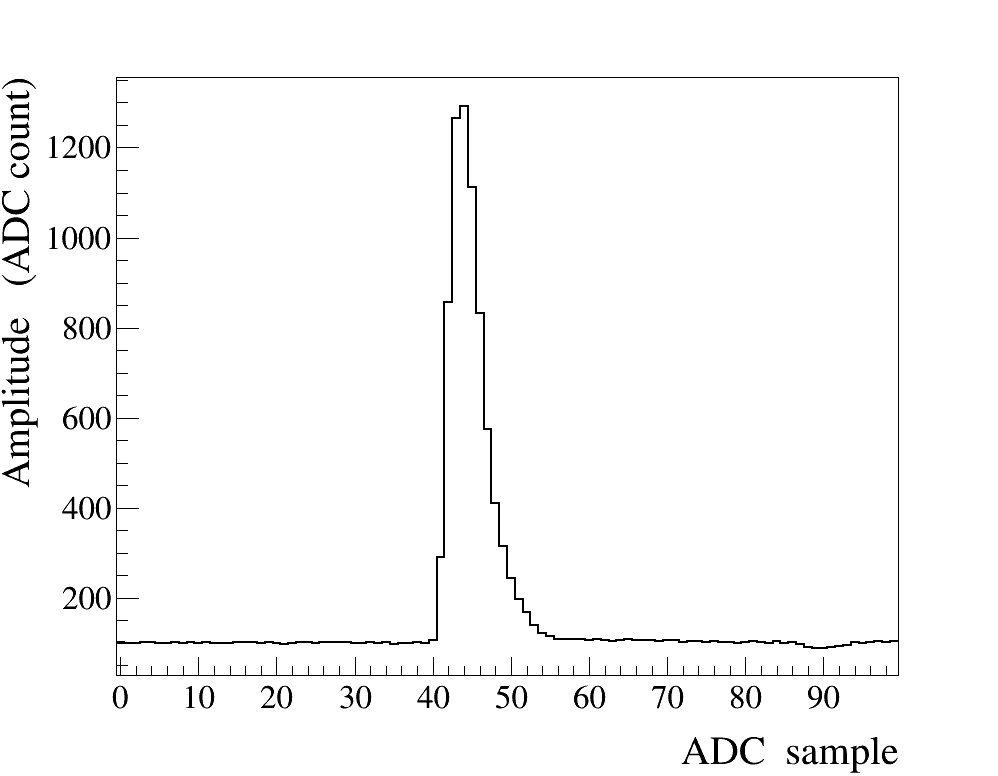}
\includegraphics[width=0.5\linewidth]{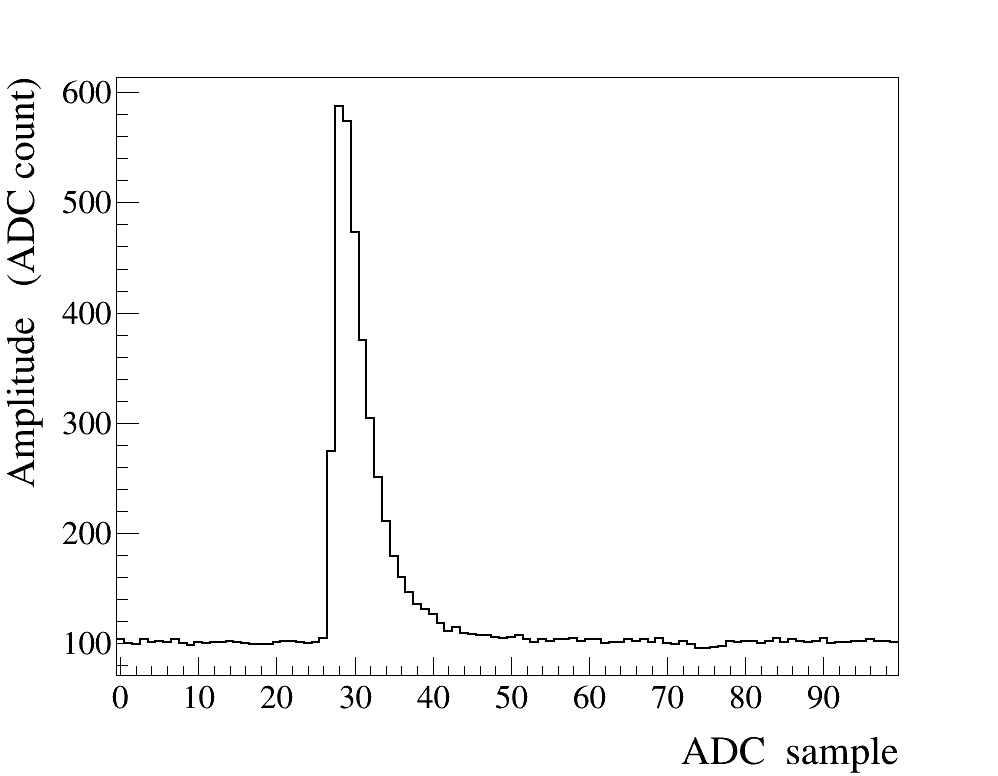}
\end{tabular}
\end{center}
\caption{Signal waveform digitized by a flash ADC, induced by the LMS (left) and by scintillation light in the PbWO$_4$ crystal (right).} 
\label{fig:ecal_pulse_shape} 
\end{figure*}

To generate a short pulse required for driving the LEDs, the input signal from the external pulse generator is split into two paths. One path is delayed using a digital delay line, implemented with Schmitt-trigger buffer gates. The delayed signal is then inverted, and both signals are combined using logic gates to produce a well-defined output pulse. This process is denoted as Trigger Pulse Shaping on Fig.~\ref{fig:lms_driver}. The shaped pulse is passed through power buffer stages to increase the current drive capability, and then fed to each individual LED driver. The final shape of the LED pulse is determined by the R6, R7, C3, and L1 components of the driver circuit and the width of the shaped trigger pulse. The driver parameters were adjusted to make an LED-induced signal pulse shape be similar to the pulse shape produced by the scintillation and Cherenkov light from the lead tungstate crystal. Signal waveforms digitized by a flash ADC for both the LMS and scintillation light are presented in the left and right plots of Fig.~\ref{fig:ecal_pulse_shape}, respectively.


The flash intensity is controlled by adjusting the regulated bias voltage, which has a maximum value of 30 V. The LED driver circuit also requires a +5 V power supply. Both voltages are provided by an MPOD 8030 power supply module placed in a Wiener crate.

\subsection{Light distribution to ECAL modules}

The homogenized light produced inside the sphere is then distributed to each individual ECAL module using acrylic optical fibers manufactured by Mitsubishi, each with a core diameter of approximately $500\;\muup{\rm m}$. The fiber core is protected by a polyethylene jacket. In order to attach optical fibers to the integrating sphere, we used a plastic plate with 55 drilled holes in it. Thirty fibers were inserted through each hole and glued together, as well as to the plate itself, using a Dymax UV-cured adhesive to ensure proper alignment and mechanical stability. After installing a total of 1650 fibers, including spares, the plate was machined to remove access length of fibers protruded through the plate and then polished to achieve optically clear fiber surface. The plate was subsequently glued to the output port of the integrating sphere, which has an effective diameter of 50.8 mm.

\begin{figure}[hbt]
\begin{center}
\includegraphics[width=0.9\linewidth,angle=0]{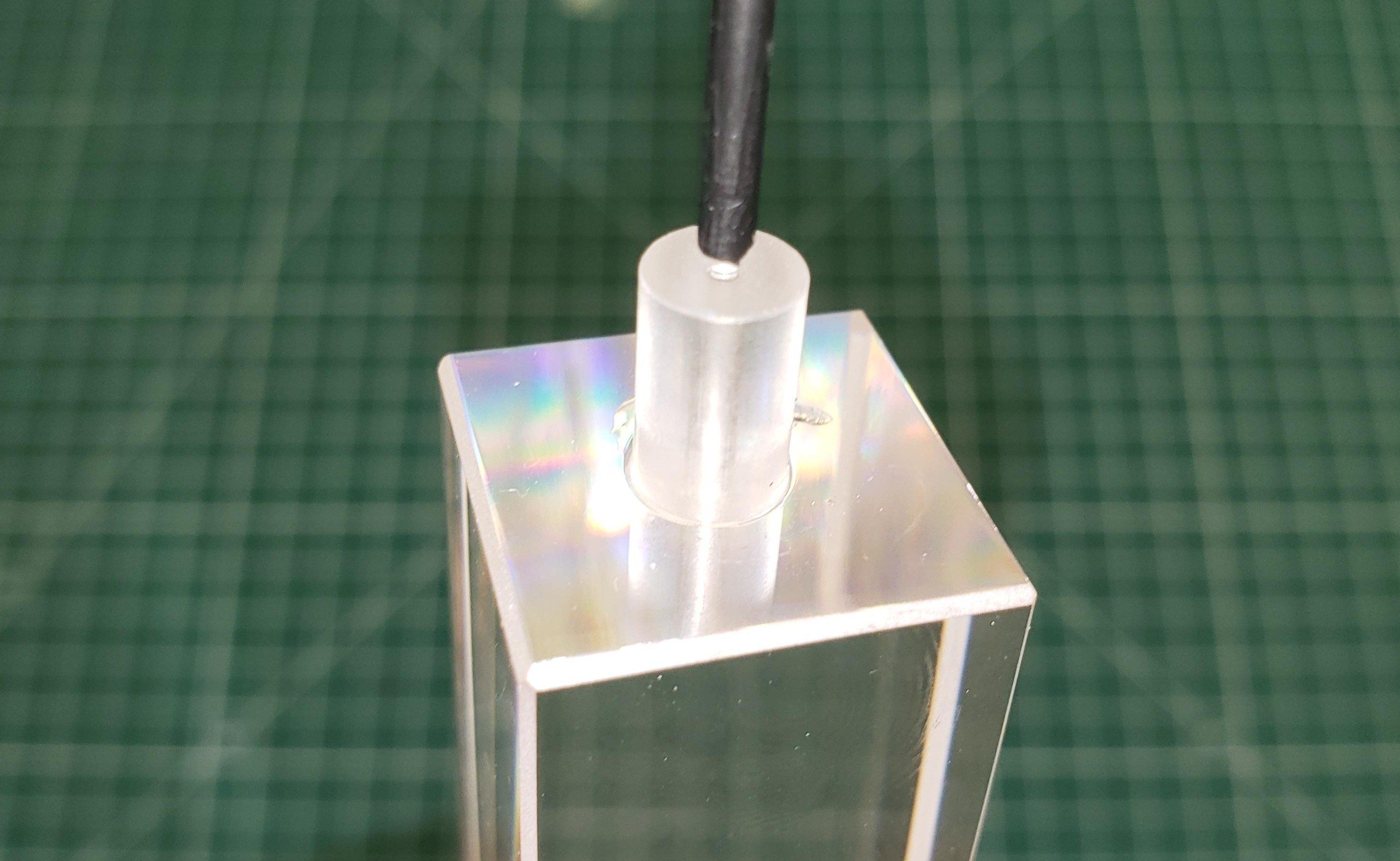}
\end{center}
\caption{The plastic cap with the inserted optical fiber glued to the face of the PbWO$_4$ crystal.}
\label{fig:lms_glued_cap}
\end{figure}

\begin{figure}[hbt]
\begin{center} 
\includegraphics[width=1.0\linewidth]{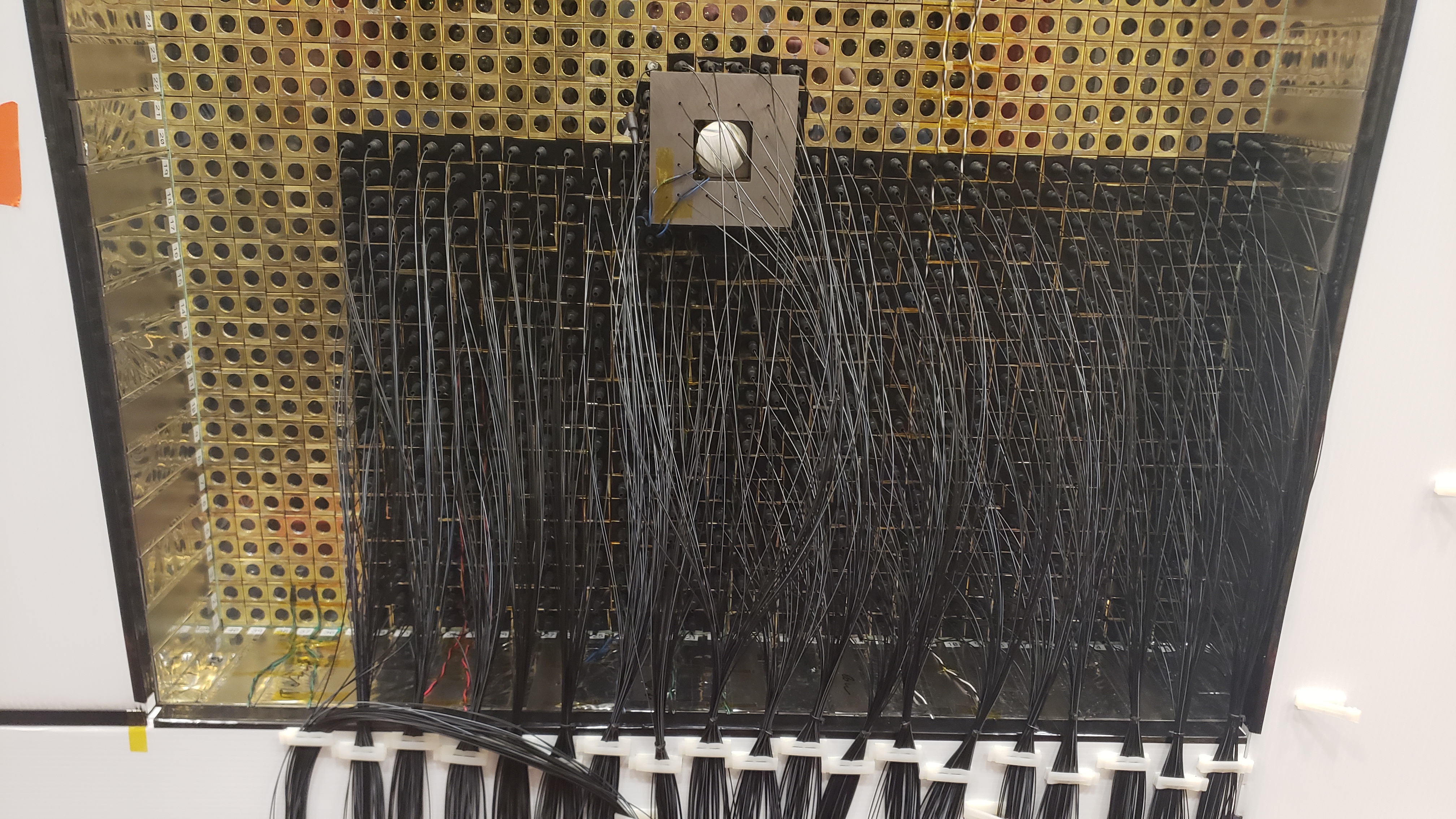}
\end{center}
\caption{Installation of optical fibers on the face of the ECAL crystals. The inner most layer of the detector surrounding the beam hole is covered by a tungsten absorber. }
\label{fig:lms_installation}
\end{figure}

On the detector side, each optical fiber is coupled to the front face of a PbWO$_4$ crystal. The coupling is performed using a custom plastic cap with a diameter of 5 mm, featuring a central hole to accommodate the fiber. After stripping the protective jacket from the fiber, it is inserted into the cap and secured with UV-cured adhesive. The cap surface is then  machined and polished before being glued directly to the crystal face. The assembly is illustrated in Fig.~\ref{fig:lms_glued_cap}.  In the actual detector module, a brass flange is installed on the face of the crystal. This flange, along with a second flange on the opposite end of the module, is connected by brass straps that provide mechanical support to hold the module components together. The front flange has a hole in it to allow the optical cap with the fiber to be attached directly to the crystal. To prevent optical cross-talk between adjacent modules, each glued cap is covered with a light-tight, pre-shaped heat-shrink sleeve and further sealed with black tape for additional light insulation. The installation of optical fibers in the detector is shown in Fig.~\ref{fig:lms_installation}. The innermost ring of the detector surrounding the beam pipe is covered by a 6 cm-thick tungsten absorber to  protect the PbWO$_4$ crystals from the high rate of electromagnetic background. To allow access to the front face of the crystals, the absorber has holes for optical fibers. The optical fibers were passed through these holes and glued to the modules prior to absorber installation.

\begin{figure}[hbt]
\begin{center}
\includegraphics[width=0.9\linewidth]{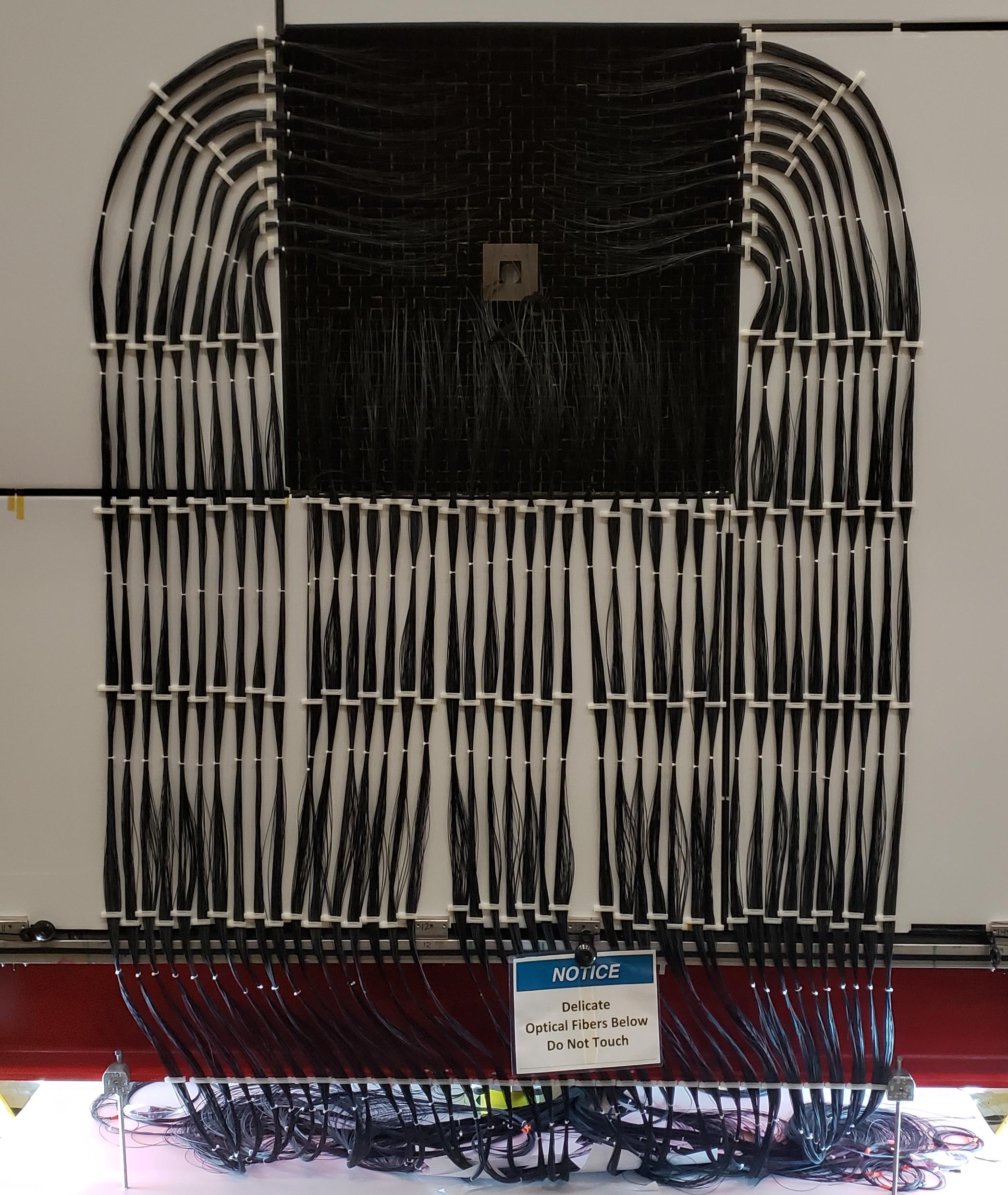}
\end{center}
\caption{Organization of the installed optical fibers on the face of the detector.}
\label{fig:lms_installed}
\end{figure}

The integrating sphere is installed inside the detector dark room, positioned at the bottom of the thermally insulated detector frame, where a stable temperature is maintained. From the dark room, the optical fibers are routed to the front face of the detector and are organized into bundles to facilitate the installation process. These bundles are attached to plastic plates that cover the lead glass modules surrounding the ECAL. Two standard optical fiber lengths were used: 2.4 meters and 4.8 meters. The shorter fibers were routed to the ECAL modules located in the lower half of the detector, while the longer fibers were used for modules in the upper region. The final installation of the optical fibers on the ECAL is shown in Fig.~\ref{fig:lms_installed}.

\subsection{Reference photo multipliers}
\begin{figure*}[t]
\begin{center}
\includegraphics[width=0.95\linewidth]{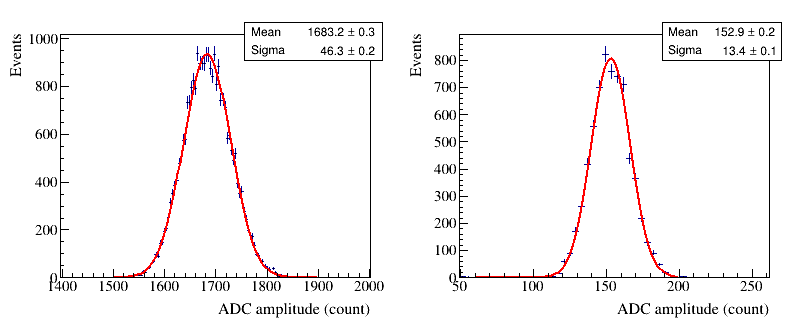}
\end{center}
\caption{Distributions of signal pulse amplitudes from the reference PMT, produced by the LED light (left) and the $\alpha$-source unit (right). The solid lines represent fits to Gaussian functions.}
\label{fig:amp_lms_alpha}
\end{figure*}

The stability of the LMS system is monitored using two reference Hamamatsu R4125 PMTs, which are installed in close proximity to the integrating sphere.  Each PMT is equipped with a light guide and housed inside a soft iron housing with additional mu-metal magnetic shielding to avoid the effects of the ambient magnetic field from the fringe field of the Solenoid magnet (which is less than 20 Gauss). Each reference PMT receives light from the LMS optical fiber and a YAP:CE pulse unit which are both attached to the face of the light guide. The unit consists of a YAP:CE scintillating crystal with a diameter of 3 mm and thickness of 1.5 mm, which is activated by an embedded ${}^{241}{\rm Am}$ $\alpha$ source. In the text, we will refer to this light source as an $\alpha$-source unit. Under nominal operating conditions, the signal pulses induced by the LMS were significantly larger than those produced by the $\alpha$-source unit. To ensure that both signals remained within the 2 V input range of the flash ADC, a thin neutral density filter foil (provided by Kodak) was used to attenuate the light from the LMS fibers by approximately a factor of two. Distributions of signal pulse amplitudes digitized by the flash ADC and induced by the LMS and the $\alpha$-source unit are shown in the left and right plots of Fig.~\ref{fig:amp_lms_alpha}, respectively. These distributions are fit with a Gaussian function, and the fit results are superimposed on the plots. High voltage for the PMTs is supplied by a CAEN 7030N module installed in a CAEN HV SY4527 mainframe. The same type of high-voltage module is used to power the PMTs throughout the entire detector. The high voltages for the reference PMTs were set to approximately 1050 V.


\section{Integration into the GlueX trigger}
\label{sec:lms_trigger}

The light monitoring system is controlled using a CAEN V1495  general-purpose VME module. This module generates triggers at a programmable rate, which are sent to the LED driver described in Section~\ref{sec:lms_driver}, to initiate light pulses. These trigger pulses are also distributed to the front panel of the Trigger Supervisor (TS) module positioned in the GlueX trigger crate, a special-purpose component that serves as a core of the GlueX trigger system~\cite{l1_trigger}. The TS produces a dedicated trigger type for the ECAL LED events, in addition to other trigger types used in the experiment. Data from the ECAL is read out using one hundred flash ADC modules installed across seven VXS crates. 

\begin{figure}[!b]
\begin{center}
\includegraphics[width=1.0\linewidth,angle=0]{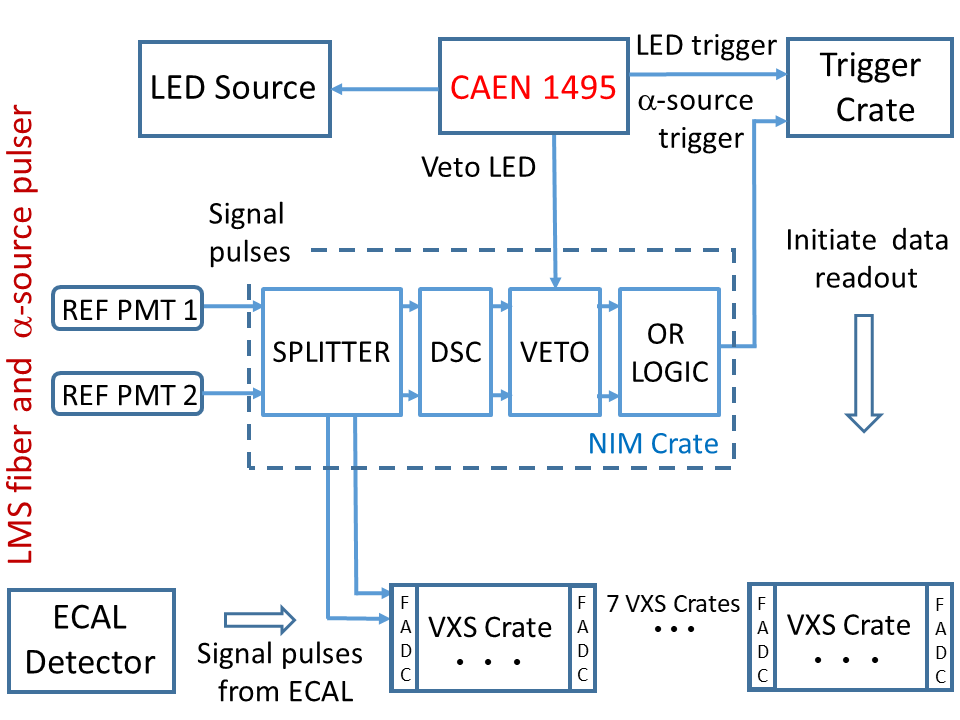}
\end{center}
\caption{Schematic view of the trigger organization in the light monitoring system.}
\label{fig:lms_trigger}
\end{figure}

In addition to the LED trigger, a special trigger type was implemented for the $\alpha$-source unit. Since both the LED optical fiber and the $\alpha$-source unit are connected to the reference PMTs, a dedicated trigger logic was implemented using NIM-based modules to distinguish between them. Signal pulses from each reference PMT are split into two paths. One path is connected directly to the flash ADC for readout, while the other is used to form the trigger for the $\alpha$-source unit. The signal is first sent to a leading-edge discriminator (DSC) and then forwarded to a logic unit, which also receives LED trigger pulses from the CAEN V1495 module. A veto logic is applied to suppress the LED pulses and select only those induced by the $\alpha$-source unit. The resulting signals from the two PMTs are combined using OR logic and sent to the front panel of the TS to form the $\alpha$-source unit trigger. Trigger timing offsets were adjusted in the CAEN V1495 module to ensure that both the LED- and $\alpha$-source-induced pulses fall within the flash ADC readout window. A schematic view of the LMS trigger logic is shown in Fig.~\ref{fig:lms_trigger}. During data acquisition, the LED system typically operates at a rate of 10 HZ, while the $\alpha$-source unit trigger rate is approximately 100 Hz.

The overall control of the LMS system, including configuring 
the trigger rate, adjusting LED flash amplitudes, and setting voltages 
on the reference PMTs, is performed using the Experimental Physics and 
Industrial Control System (EPICS)~\cite{epics}. EPICS provides a user interface that  
allows to define required parameters and communicates these 
settings to the corresponding electronic modules.

\section{Performance}
\label{sec:lms_performance}

The main components of the LMS, including the light source and optical fiber, were installed after all calorimeter modules had been placed in the experimental hall. The LMS was extensively used during the installation of PMT dividers. Signal waveforms induced by the LMS were evaluated after the dividers were mounted on each detector layer, helping to identify and resolve various issues related to electronics and cabling.

\begin{figure}[t]
\begin{center}
\includegraphics[width=1.0\linewidth]{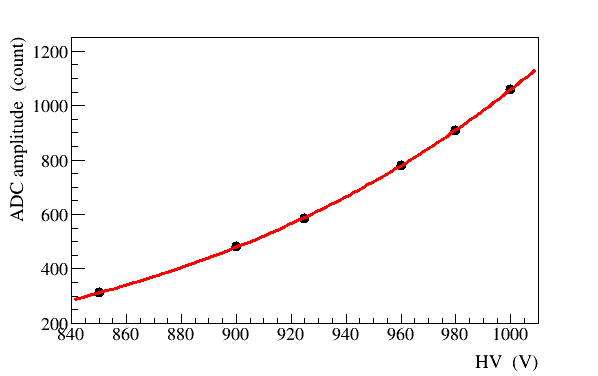}
\end{center}
\caption{Dependence of the flash ADC amplitude on the PMT high voltage. Superimposed on the plot is a fit to the power law function. }
\label{fig:ecal_calib_curve}
\end{figure}

The LMS was also used during detector commissioning with both cosmic ray and beam data. The operating high voltages of the detector PMTs were set based on LMS calibration curves to ensure a uniform response across all channels. This was especially critical for the GlueX trigger, which is based on energy deposition in the calorimeters and requires that the flash ADC amplitudes in the ECAL are equalized for a given energy of the incoming photon. An example of such a calibration curve obtained using the LMS is shown in Fig.~\ref{fig:ecal_calib_curve}, illustrating the typical dependence of the signal pulse amplitude in an ECAL module on the applied high voltage. A power-law function is superimposed on the plot to fit the data, demonstrating the gain behavior of the PMTs. The typical gain exponent extracted from the fit is approximately 7.5.

The LMS was continuously operated during data taking, enabling real-time monitoring of the detector's stability. The typical bias voltage used to control the LED flash amplitudes was set to 20 V, with a maximum value of 30 V. These settings resulted in average pulse amplitudes in the ECAL modules of approximately 1000 flash ADC counts. This amplitude corresponds to the energy-equivalent response in the central shower module for an incoming photon with an energy of 2.8 GeV. The average spread in amplitude across the calorimeter modules was about $30\%$.

\begin{figure}[t]
\begin{center}
\includegraphics[width=1.0\linewidth]{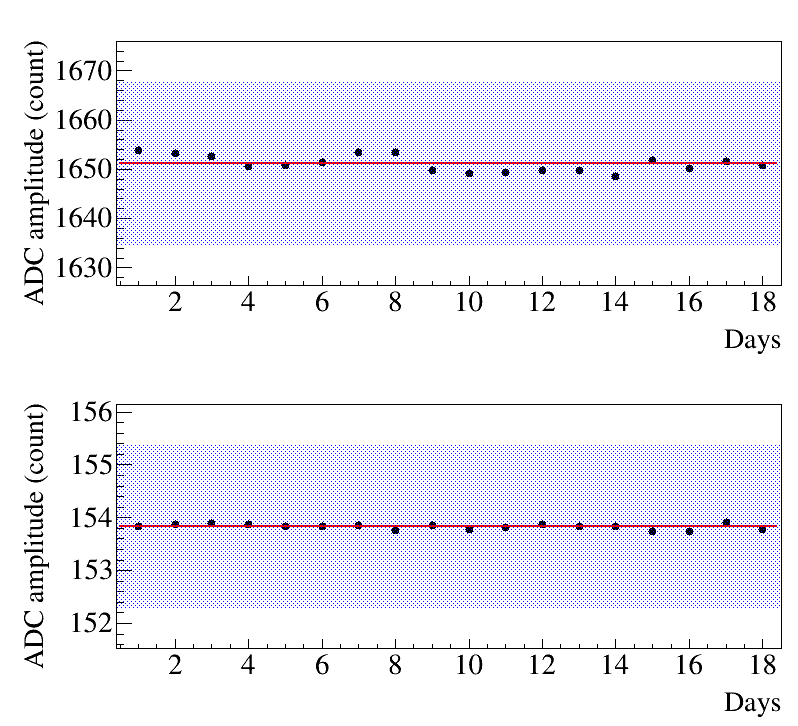}
\end{center}
\caption{Signal pulse amplitudes induced by the LED light (top) and $\alpha$-source unit (bottom) for two reference PMTs as a function of time.}
\label{fig:ref_stab}
\end{figure}

The stability of the light monitoring system itself was evaluated using reference PMTs, which were calibrated with respect to the $\alpha$-source unit. As an example, the signal pulse amplitudes recorded by a reference PMT over a period of about two weeks are shown in Fig.~\ref{fig:ref_stab}.  The amplitudes induced by the LED light and the $\alpha$-source unit are presented in the top and bottom plots, respectively. The solid line in each plot represents the average amplitude over the monitoring period, while the hatched region indicates a $\pm 1\%$ deviation from this mean. As shown, the amplitudes from both the LED and $\alpha$-source unit remained stable at the sub-percent level during the monitoring period. To account for possible instabilities in the LED source, the LED light intensity delivered to the ECAL modules is corrected based on the ratio of the amplitudes from the LED and the $\alpha$-source unit, as measured by the reference PMT.

\begin{figure}[t]
\begin{center}
\includegraphics[width=1.0\linewidth]{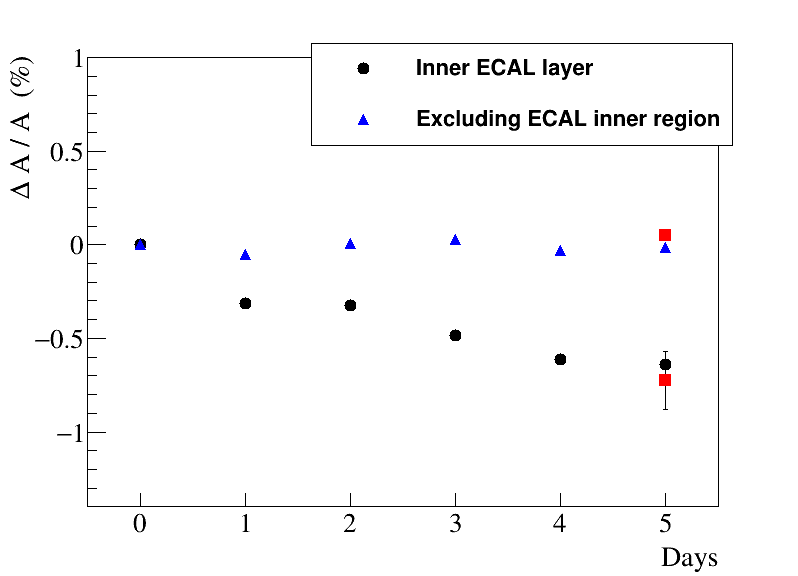}
\end{center}
\caption{The relative change in amplitude induced by the LMS during five consecutive days of data taking. Box markers represent the relative PMT gain corrections for the innermost ECAL layer near the beamline and for the ECAL modules outside the inner region, as obtained from $\pi^0$ calibration.}
\label{fig:amp_stab_time}
\end{figure}

In most calorimeter modules the LMS amplitudes remain relatively stable over time. However, a slight degradation is observed in detector modules located near the beam pipe. These modules are exposed to a high flux of particles produced by electromagnetic interactions in the GlueX detector target and the surrounding beamline materials. The typical hit rate in these modules reaches approximately 500 kHz at a threshold of about 5 mV. Figure~\ref{fig:amp_stab_time} shows an example of LMS amplitude stability over time. The vertical axis displays the relative change in amplitude, defined as $\Delta A / A = (A_{t} - A_{0}) / A_{0}$, where $A_t$ is the amplitude measured at a given time $t$, and $A_0$ is the amplitude recorded during the initial measurement. The triangle markers represent the relative amplitudes for an ECAL module located in the central region of the detector, which reflects the typical behavior observed in most modules. In contrast, the black circles illustrate the amplitude evolution for a module in the innermost detector layer near the beamline, which is not shielded by a tungsten absorber. In this case, a small degradation in amplitude is observed. This relative decrease in LMS amplitude can be independently confirmed by measuring changes in the photomultiplier tube response using the $\pi^0$ calibration method. The $\pi^0$ invariant mass is reconstructed from two electromagnetic showers produced in the decay $\pi^0 \to \gamma\gamma$. A PMT amplitude correction, referred to here as a gain correction, is applied to account for any shift in the reconstructed $\pi^0$ mass resulting from a decrease in PMT response to scintillation light from electromagnetic showers over time. The relative gain variations for the modules in the innermost detector layer and for the ECAL modules outside the inner region are indicated by box markers in Fig.~\ref{fig:amp_stab_time}. It is important to note that the scintillation light yield in PbWO$_4$ crystals depends on temperature, in contrast to the LED light injected by the LMS. However, during the run period, the temperature was controlled to better than $0.1\%$, so temperature-dependent effects are expected to be small. 

\begin{figure}[hbt]
\begin{center}
\includegraphics[width=1.0\linewidth]{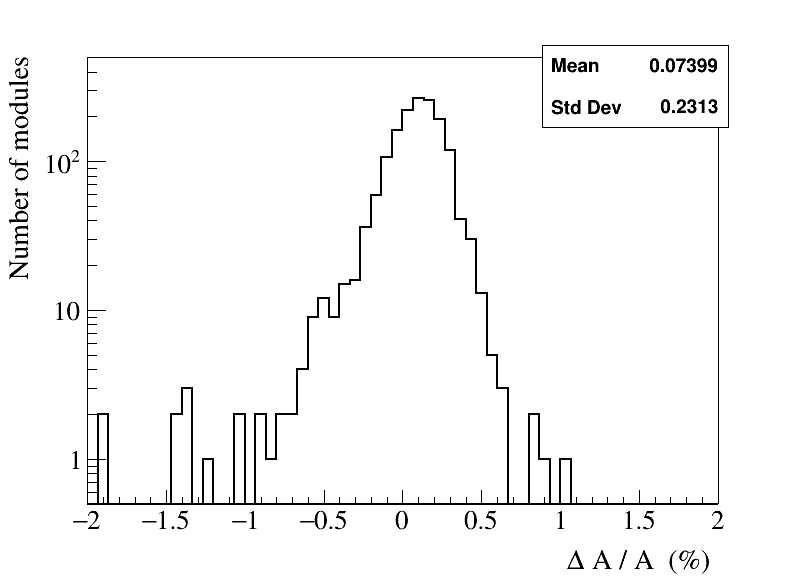}
\end{center}
\caption{Relative difference in LMS-induced signal pulse amplitudes of all ECAL modules after five days of 
continuous data taking.}
\label{fig:amp_stab}
\end{figure}

The overall relative change in signal pulse amplitudes due to the LMS across all ECAL modules over five days of data taking is summarized in Fig.~\ref{fig:amp_stab}. The distribution is slightly asymmetric, with a small enhancement in the region of decreased amplitudes (negative $\Delta A/A$), mainly due to the modules near the beam pipe. The overall variation in the relative amplitude change is  small, with a standard deviation of just $0.23\%$.


\section{Summary}
\label{sec_summary}
The light monitoring system  was designed for the lead tungstate calorimeter in Hall D at Jefferson Lab.  It is based on multiple blue LEDs, with light from the source individually distributed to 1596 calorimeter modules. The LMS was fabricated, installed on the detector, and successfully used during the electromagnetic calorimeter commissioning. It was continuously operated throughout the first physics run with the ECAL detector in 2025. The LMS enables monitoring of signal amplitudes in the ECAL modules induced by the LMS light and will be used to support PMT gain calibration. System stability was monitored using a reference PMT. The LMS demonstrated stable and reliable performance during the data production run.

\section{Acknowledgments}

This material is based upon work supported by the U.S. Department of Energy, Office of Science, Office of Nuclear Physics under contract DE-AC05-06OR23177, as well as NSF grants PHY-1812396, PHY-2111181, and PHY-2412800.  We are grateful to the Hall D technical group for their assistance during the installation of the light monitoring system.

\bibliographystyle{elsarticle-num}

\bibliography{ecal_lms}

\begin{thebibliography}{1}
\expandafter\ifx\csname url\endcsname\relax
  \def\url#1{\texttt{#1}}\fi
\expandafter\ifx\csname urlprefix\endcsname\relax\def\urlprefix{URL }\fi
\expandafter\ifx\csname href\endcsname\relax
  \def\href#1#2{#2} \def\path#1{#1}\fi

\bibitem{gluex_det}
S.~Adhikari, et~al., {The GLUEX beamline and detector}, Nucl. Instrum. Meth. A
  987 (2021) 164807.
\newblock \href {http://arxiv.org/abs/2005.14272} {\path{arXiv:2005.14272}},
  \href {http://dx.doi.org/10.1016/j.nima.2020.164807}
  {\path{doi:10.1016/j.nima.2020.164807}}.

\bibitem{jef}
M.~Dugger, et~al., Eta decays with emphasis on rare neutral modes, Tech. Rep.
  GlueX-Doc-2460, JLab, available at:
  \url{https://www.jlab.org/exp_prog/proposals/14/PR12-14-004.pdf} (2014).

\bibitem{ecal}
A.~Somov, {Lead tungstate calorimeter of the Jefferson Lab Eta Factory
  experiment}, EPJ Web Conf. 320 (2025) 00058.
\newblock \href {http://dx.doi.org/10.1051/epjconf/202532000058}
  {\path{doi:10.1051/epjconf/202532000058}}.

\bibitem{fadc250}
F.~Barbosa, et~al., {A VME64x, 16-Channel, Pipelined 250 MSPS Flash ADC With
  Switched Serial (VXS) Extension}, Tech. rep., Jefferson Lab,
  \href{https://halldweb.jlab.org/doc-public/DocDB/ShowDocument?docid=1022}{Technical
  Report GlueX-doc-1022} (hyperlink) (Apr. 2007).

\bibitem{fcal2}
A.~Asaturyan, et~al., {Electromagnetic calorimeters based on scintillating lead
  tungstate crystals for experiments at Jefferson Lab}, Nucl. Instrum. Meth. A
  1013 (2021) 165683.
\newblock \href {http://dx.doi.org/10.1016/j.nima.2021.165683}
  {\path{doi:10.1016/j.nima.2021.165683}}.

\bibitem{l1_trigger}
A.~Somov, {Development of level-1 triggers for experiments at Jefferson Lab},
  AIP Conf. Proc. 1560~(1) (2013) 700--702.
\newblock \href {http://dx.doi.org/10.1063/1.4826876}
  {\path{doi:10.1063/1.4826876}}.

\bibitem{epics}
L.~R. Dalesio, et~al., {The Experimental Physics and Industrial Control System
  architecture: Past, present, and future}, Nucl. Instrum. Meth. A 352 (1994)
  179--184.
\newblock \href {http://dx.doi.org/10.1016/0168-9002(94)91493-1}
  {\path{doi:10.1016/0168-9002(94)91493-1}}.

\end{thebibliography}

\end{document}